\documentclass[aps,prl,amsmath,amsfonts,amssymb,twocolumn,showpacs]{revtex4-1}
\usepackage[latin1]{inputenc}
\usepackage{graphicx}
\usepackage[english]{babel}
\usepackage[usenames,dvipsnames]{color}

\usepackage{amsfonts}
\usepackage{amsmath}
\usepackage{amssymb}
\usepackage{bibunits}

\usepackage{soul}

\newcommand{\beq}{\begin{equation}}
\newcommand{\eeq}{\end{equation}}
\newcommand{\nn}{\nonumber}
\newcommand{\ket}[1]{|#1\rangle}
\newcommand{\bra}[1]{\langle #1|}

\makeatletter
\@ifundefined{textcolor}{}
{%
 \definecolor{BLACK}{gray}{0}
 \definecolor{WHITE}{gray}{1}
 \definecolor{RED}{rgb}{1,0,0}
 \definecolor{GREEN}{rgb}{0,1,0}
 \definecolor{BLUE}{rgb}{0,0,1}
 \definecolor{CYAN}{cmyk}{1,0,0,0}
 \definecolor{MAGENTA}{cmyk}{0,1,0,0}
 \definecolor{YELLOW}{cmyk}{0,0,1,0}
 }

\defaultbibliography{QuPiston_ArXiv}
\defaultbibliographystyle{apsrev4-1}

\begin{document}

\title{Quantum work by a single photon}

\author{D. Valente$^{1,2}$}
\email{valente.daniel@gmail.com}

\author{F. Brito$^{3}$}

\author{R. Ferreira$^{2}$}

\author{T. Werlang$^{1}$}

\affiliation{$^{1}$ 
Instituto de F\'isica, Universidade Federal de Mato Grosso, CEP 78060-900, Cuiab\'a, MT, Brazil
}

\affiliation{
$^{2}$ 
Laboratoire Pierre Aigrain, \'Ecole Normale Sup\'erieure - Paris, Centre National de la Recherche Scientifique, 24 rue Lhomond, F-75005 Paris, France
}

\affiliation{
$^{3}$ 
Instituto de F\'isica de S\~ao Carlos, Universidade de S\~ao Paulo, C.P. 369, 13560-970, S\~ao Carlos, SP, Brazil
}

\begin{abstract}
The work performed by a classical electromagnetic field on a quantum dipole is well known in quantum optics.
The absorbed power linearly depends on the time derivative of the average dipole moment, in that case.
The following problem, however, still lacks an answer:
can the most elementary electromagnetic pulse, consisting of a single-photon state, perform work on a quantum dipole?
As a matter of fact, the average quantum dipole moment exactly vanishes in such a scenario.
In this paper, we present a method that positively answers to this question, by combining techniques from the fields of quantum machines and open quantum systems.
Quantum work here is defined as the unitary contribution to the energy variation of the quantum dipole.
We show that this quantum work corresponds to the energy spent by the photon pulse to dynamically Stark shift the dipole.
The non-unitary contribution to the dipole energy is defined here as a generalized quantum heat.
We show that this generalized quantum heat is the energy corresponding to out-of-equilibrium photon absorption and emission.
Finally, we reveal connexions between the quantum work and the generalized quantum heat transferred by a single photon and those by a low-intensity coherent field.

\end{abstract}


\maketitle
\begin{bibunit}
{\bf Electromagnetic work on an electric dipole.--}
The classical electromagnetic theory predicts that a time-dependent electric field $E(t)$ is able to perform work on a rate $\dot{W} = E(t) \dot{d}(t)$ on a classical electric dipole of moment $d(t)$.
When a laser interacts with an atom, quantum mechanical effects may play a relevant role.
A classical time-dependent electric field performs work on an atom at a rate 
$\dot{W} = E(t) \langle \dot{\hat{d}}(t)\rangle$, 
where the quantum average of the dipole moment $\langle {\hat{d}}(t)\rangle$ is the relevant parameter \cite{cohen}.
The laser can be described quantum mechanically, as well. 
A quantized electromagnetic field $\hat{E}$ initially prepared in a coherent quantum state $\ket{\alpha(0)}$ induces the same dynamics on the average dipole as does a classical field $E(t)$, as long as the classical field value matches the average value of the quantum field, $E(t) = \bra{\alpha(t)} \hat{E} \ket{\alpha(t)}$.

The non-classical nature of the atom-field interaction becomes more evident when the atom is driven by the most elementary electromagnetic pulse, prepared in the single-photon state described as
\beq
\ket{1} = \sum_{\omega} \phi_{\omega}(0) a_\omega^\dagger \ket{0}.
\label{ket1}
\eeq$\ket{0} = \prod_\omega \ket{0_\omega}$ is the vacuum state of the field in each mode of frequency $\omega$ with creation operator $a_\omega^\dagger$, forming a continuum $\sum_\omega \rightarrow \int d\omega \varrho_0$, with flat spectral density $\varrho_0$.
The probability amplitude for the initial occupation of each mode is $\phi_{\omega}(0)$, which is normalized, $\sum_\omega |\phi_\omega(0)|^2 = 1$.
This electromagnetic pulse contains exactly a single excitation, $ \sum_\omega a^\dagger_\omega a_\omega \ket{1} = \ket{1}$.
In the single-photon state, the average electric field equals to zero,
$
\bra{1} \hat{E} \ket{1} = 0,
$
where
$\hat{E} = \sum_\omega i \epsilon_\omega (a_{\omega} e^{ik_{\omega} z} - a^\dagger_{\omega} e^{-ik_{\omega} z})$ 
\cite{glauber63,domokos}.
Let us further assume that the global state of the atom-field system is $\ket{\xi(t)} = \psi(t) \ket{e,0} + \sum_{\omega} \phi_{\omega}(t) a_\omega^\dagger \ket{g,0}$, as discussed below.
$\ket{e}$ ($\ket{g}$) is the excited (ground) state of the two-level system (TLS).
It follows from $\ket{\xi(t)}$ that the average dipole of an atom driven by the single-photon pulse also equals to zero,
i.e., $\langle \hat{d}(t) \rangle = \mbox{Tr}_s[\rho_s(t) \hat{d}]  = 0$, 
where $\hat{d} = d_{eg} (\ket{e}\bra{g} +\ket{g}\bra{e})$.
The dipole vanishes here because $\bra{e} \rho_s(t) \ket{g} = 0$, where the reduced state of the TLS is $\rho_s(t) = \mbox{Tr}_{\mathrm{field}}[\ket{\xi(t)}\bra{\xi(t)}]$, obtained by tracing out the field degrees of freedom.

In this paper, we address the following problem. 
In that single-photon scenario, of vanishing average field, 
$\bra{1} \hat{E} \ket{1} = 0$, 
and vanishing average dipole, 
$\langle \hat{d}(t) \rangle = 0$,
would the electromagnetic field still be capable of performing work on a quantum electric dipole?
A proper definition of work is key.
We combine techniques from out-of-equilibrium open quantum systems and from quantum machines to tackle the problem using a fully-quantum model.
The quantum work analyzed here gives the amount of energy a single photon can transfer to the TLS dipole by means of a unitary contribution, as far as the reduced TLS dynamics is concerned.
We show that this quantum work corresponds to the energy spent by the photon to dynamically Stark shift the TLS energy levels.
We establish connexions with a semiclassical scenario, within the low-excitation regimes of both the quantum and the semiclassical models.
Namely, under appropriate conditions, the quantum work by a single photon equals the reactive energy, related with the real part of the linear susceptibility $\chi'(\omega)$, reversibly stored in a quantum dipole from a low-intensity coherent pulse.
Also, under the same conditions, the non-unitary contribution to the dipole energy absorbed from the photon equals the absorptive contribution from a low-intensity coherent pulse, related with the imaginary part $\chi''(\omega)$.
As discussed below, this non-unitary contribution can be identified with a generalized quantum heat.

Our results are relevant not only to the field of light-matter interaction at the single-photon level \cite{domokos,fan05,chang07,sand09,jmg10,tsai10,scully10,sand12,dvNJP,yeo13,ck,wallraff,singlephoton},
but also to quantum thermodynamics and quantum autonomous machines \cite{mahler05,mahler08,QA0,QA1,QA3,QA4,QAKurizkiAlicki,QA2,QA50,QA5,QA6,QA7}.
The quantum nature of the electromagnetic pulse here presents features of a quantized piston in an autonomous machine.
Also, the single-photon pulse provides a non-Markovian out-of-equilibrium quantum environment for the TLS, which involves an extension of quantum thermodynamics to situations out of thermal equilibrium \cite{PRL04,nature06,PRL08,PRL09,PRB12,PhotonOutThEq, NMQuTh}.

{\bf Quantum work.--}  A classically-driven quantum system was studied in the late seventies \cite{pusz78, alicki79}.
The case where the quantum system was both classically driven and weakly coupled to two Markovian thermal reservoirs at distinct temperatures was further investigated, where the Carnot bound was found for the efficiency of the quantum engine \cite{alicki79}.
The physical principle behind quantum work was a quantum version of the first law of thermodynamics.
The internal system's energy corresponded to the average energy $U(t) = \mbox{Tr}[\rho_s(t) H_s(t)]$, where $\rho_s(t)$ is the system's quantum state.
$H_s(t)$ is the time-dependent hamiltonian appearing in the unitary part of the master equation for ${\rho}_s(t)$.
The variation of the internal energy was decomposed as $d{U} = \dj{W}_A + \dj{Q}_A$. 
$\dj{W}_A = \mbox{Tr}[\rho_s(t) \partial_t {H}_s(t)] dt$ was the unitary contribution to the system's energy variation. 
It vanishes if the external conditions do not change in time, i.e. if $\partial_t{H}_s(t) = 0$.
The non-unitary contribution was given by $\dj{Q}_A = \mbox{Tr}[\partial_t{\rho}_s(t) {H}_s(t)] dt$. 
It vanishes for a purely unitary system's dynamics generated by ${H}_s(t)$. Taking the quantum dynamics of the drive itself fully into account is a central problem in quantum thermodynamics.
To that end, generalized concepts of quantum work and quantum heat have been proposed for autonomous quantum machines  \cite{mahler05,mahler08,QA0,QA1,QA3,QA4}.
An autonomous machine is that whose global Hamiltonian, of the complete system, is time independent \cite{classmech19351,classmech19352}.
Ref.\cite{mahler08}, for instance, considers the local von Neumann entropy of a given subsystem as the fundamental concept for defining generalized quantum work and heat exchanges within autonomous machines.
A complete relationship between this generalized quantum heat and thermodynamic heat flow at the single-quantum level \cite{PRL04,nature06,PRL08,PRL09,PRB12} remains an open problem.
In these quantum driving scenarios \cite{mahler08,QA0,QA1,QA3,QA4}, knowing the master equation for the reduced dynamics of the subsystem of interest is also necessary.

In our case, the master equation for the reduced dynamics of the TLS driven by a single-photon pulse can be derived from only one assumption:
the global field-TLS quantum state is written as $\ket{\xi(t)}$, defined after Eq.(\ref{ket1}).
This means that the global state remains pure throughout the entire time evolution and that the total number of excitations in the system is conserved.
The exact master equation we employ here has been derived in Refs.\cite{buzek00,dv16, dv17}, which reads
\beq
\partial_t \rho_s(t) = -\frac{i}{\hbar} [H_s(t),\rho_s(t)] + \mathcal{L}_t \{ \rho_s(t) \}.
\label{ME}
\eeq
The effective TLS Hamiltonian in the above equation is
\beq
H_s(t) = \hbar \omega_s(t)\ \sigma_+ \sigma_-,
\label{Hst}
\eeq
where $\sigma_+ = \sigma_-^\dagger = \ket{e} \bra{g}$.
The non-unitary part is
\beq
\mathcal{L}_t \{ \rho_s(t) \} 
= 
\Gamma(t) \left( \sigma_- \rho_s(t) \sigma_+ - \frac{1}{2} \{  \sigma_+ \sigma_-, \rho_s(t) \}  \right),
\label{NU}\eeq
where $\{.,.\} $ is the anticommutator.
$\mathcal{L}_t \{ \rho_s(t) \}$ happens to be in the Lindblad form, even though the field is treated here non perturbatively, i.e., the quantum dynamics of the field is fully described within $\ket{\xi(t)}$.
In Eq.(\ref{Hst}), the TLS time-dependent frequency induced by the single-photon packet is defined as
$
\omega_s(t) \equiv -\mbox{Im}[\dot{\psi}(t)/\psi(t)]
$.
The physical origin of the time dependency in $\omega_s(t)$ is a dynamic Stark-shift effect, induced by the single-photon packet \cite{dv17}.
In Eq.(\ref{NU}), the time-dependent decay rate induced by the out-of-equilibrium quantized pulse is 
$
\Gamma(t) \equiv  -2\ \mbox{Re}[\dot{\psi}(t)/\psi(t)]
$.
A negative decay rate $\Gamma(t) < 0$ is a signature of the non-Markovianity of the TLS dynamics, induced by the single-photon packet \cite{dv16}.

We identify quantum work by a single-photon pulse on a TLS dipole here with the unitary contribution to the TLS energy, as inspired by Refs.\cite{alicki79,mahler08},
\begin{eqnarray}
W_1 &\equiv& \int_{t_0}^{t_f} \mbox{Tr}[\rho_s(t) \partial_t{H}_s(t)] \ dt \nn \\
&=& \int_{t_0}^{t_f} |\psi(t)|^2 \Big{(} \partial_t {\hbar \omega}_s(t) \Big{)}  dt.
\label{w1}
\end{eqnarray}
A crucial result of this paper is in the second line of Eq.(\ref{w1}).
It evidences how the dynamic Stark shift of $\omega_s(t)$, as induced by the single-photon pulse \cite{dv17}, plays a critical role in the quantum work we analyze.
The relation between $\omega_s(t)$ and the average TLS-field interaction energy is explained in the following paragraph.
We identify the generalized quantum heat here with the non-unitary contribution to the TLS energy,
\begin{eqnarray}
Q_1 &\equiv& \int_{t_0}^{t_f} \mbox{Tr}[\partial_t{\rho}_s(t) {H}_s(t)] dt  \nn \\
&=& \int_{t_0}^{t_f} \big{(} \partial_t {|\psi(t)|^2} \big{)} \hbar \omega_s(t) dt,
\label{q1}
\end{eqnarray}
also inspired in Refs.\cite{alicki79,mahler08}.
This generalized heat crucially depends on the variation of the TLS excitation probability $|\psi(t)|^2$, by means of photon absorption and emission.

It is worth stressing that all the results so far, from Eq.(\ref{ME}) to (\ref{q1}), are general, relying only on the form of the state 
$\ket{\xi(t)}$.
They show that the TLS excited-state amplitude, $\psi(t)$, is the key quantity to be evaluated.  
In order to go a step further in the explicit calculation of $\omega_s(t)$, $\Gamma(t)$, and of the generalized quantum work and heat, the expression for the global hamiltonian is needed. 
We assume a dipole TLS-field coupling, 
$H_{\mathrm{int}} = -\hat{d}\hat{E}$, where
$\hat{d} = d_{eg} (\sigma_+ + \sigma_-)$ is the TLS dipole operator and 
$\hat{E} = \sum_\omega i \epsilon_\omega (a_{\omega} e^{ik_{\omega} z_s} - a^\dagger_{\omega} e^{-ik_{\omega} z_s})$ 
is the quantized field at position $z_s$ \cite{cohen,domokos}.
The dipole and the field are assumed to be linearly polarized in the same direction.
The TLS center of mass is assumed to remain fixed at the origin of the reference frame at all times during the interaction, $z_s = 0$, since we are interested in the effect of the field on the TLS dipole only.
By defining the vacuum Rabi frequency as $g = d_{eg} \epsilon_{\omega_0}/\hbar$ and doing a rotating-wave approximation (RWA) in the interaction Hamiltonian \cite{cohen}, we obtain 
$
H_{\mathrm{int}} = -i\hbar g \sum_\omega (a_\omega \sigma_+ - a^\dagger_\omega \sigma_-)
$.
The free TLS Hamiltonian is $H_s = \hbar {\omega}_0  \sigma_+ \sigma_-$ and the free field Hamiltonian is $H_{\mathrm{field}} = \sum_{\omega} \hbar \omega a^\dagger_{\omega} a_{\omega}$, so that the global Hamiltonian is $H = H_s + H_{\mathrm{field}} + H_{\mathrm{int}}$.
Conservation of the total number of excitations is guaranteed under RWA, $[N,H]=0$, where 
$N = \sigma_+ \sigma_- + \sum_{\omega} a^\dagger_\omega a_\omega$.
The Schr\"odinger equation $i\hbar \partial_t \ket{\xi(t)} = H \ket{\xi(t)}$ is solved for the state $\ket{\xi(t)}$, defined below Eq.(\ref{ket1}).
We stress that $H$ here is itself time independent, $\partial_t {H} = 0$, assuring that the global TLS-field system is autonomous.
By formally integrating $\phi_\omega(t)$, substituting it in the equation for $\psi(t)$ and doing a Wigner-Weisskopf approximation in the continuum limit $\sum_\omega \rightarrow \int d\omega \varrho_0$ \cite{dvNJP,dv16,dv17}, we obtain
\beq
\partial_t \psi(t) = -\left( \frac{\Gamma_0}{2} + i\omega_0 \right) \psi(t) - g\ \phi(0,t).
\label{psit}
\eeq
The TLS spontaneous emission rate is $\Gamma_0 = 4\pi g^2 \varrho_0$.
The input single-photon wave packet at the TLS position, $z_{s} = 0$, is $\phi(z_s, t) = \sum_{\omega} \phi_{\omega}(0)e^{i(k_{\omega} z_{s} - \omega t)}$.
This model is not limited to low excitations of the TLS, i.e., $0 \leq |\psi(t)|^2 \leq 1$, respecting the normalization condition, 
$\bra{\xi(t)} \xi(t)\rangle = 1$.
Here, we are interested in the case where the atom is initially in state $\ket{g}$, so the excitation is completely in the field at $t=0$.
Therefore, the solution of Eq.(\ref{psit}) for $\psi(0) = 0$ reads 
$\psi(t) = - g\int_0^t e^{-(\Gamma_0/2 + i\omega_0)(t-t')}  \phi(0,t')  dt'$.
From Eq.(\ref{psit}), we get that $\omega_s(t) = \omega_0 + \delta_{\mathrm{eff}}(t)$, where 
$
\delta_{\mathrm{eff}}(t) \equiv g \mbox{Im}[\phi(0,t)/\psi(t)]
$
is the effective time-dependent modulation of the TLS frequency.
The influence of $g$ on $\omega_s(t)$ suggests a relation with the average TLS-field interaction energy.
Indeed, we find that
\beq
\hbar \delta_{\mathrm{eff}}(t) = \langle H_{\mathrm{int}} (t) \rangle / (2 |\psi(t)|^2),
\eeq
where $\langle H_{\mathrm{int}} (t) \rangle \equiv \bra{\xi(t)} H_{\mathrm{int}} \ket{\xi(t)}$.
This clarifies how the single-photon quantum work is related with the TLS-field interaction, since $W_1 = \int_{t_0}^{t_f} |\psi(t)|^2\big{(} \partial_t {\hbar \delta}_{\mathrm{eff}}(t)\big{)}  dt$. The time-dependent effective decay rate can also be obtained from Eq.(\ref{psit}), 
$\Gamma(t) = \Gamma_0 + 2g\mbox{Re}[\phi(0,t)/\psi(t)]$.
Its influence on the generalized quantum heat becomes more evident when we note that $\partial_t |\psi(t)|^2 = -\Gamma(t) |\psi(t)|^2$, which follows from the definition of $\Gamma(t)$.
Hence, $Q_1 = \int_{t_0}^{t_f} \big{(} - \Gamma(t) |\psi(t)|^2 \big{)} \hbar \omega_s(t) dt$.
Generalized heat can only be absorbed, $Q_1 > 0$, if the effective TLS decay rate is negative, $\Gamma(t) < 0$.

We choose a pulse as yielded by typical single-photon sources \cite{cohen,scully97}, 
$\phi^{(+)}(z,0) = N\Theta(-z)e^{(\Delta/2 + i \omega_L)z/c}$, 
with normalization 
$N = \sqrt{2 \pi \varrho_0 \Delta}$. 
It is prepared in the negative side of the one-dimensional reference frame, $z < 0$, being $\Theta(z)$ the Heaviside step function.
Here, $\omega_L$ is the central pulse frequency.
A TLS-field detuning is defined as $\delta_L = \omega_L - \omega_0$.
This single-photon pulse is prepared in the modes with positive wavevectors, $k_\omega = \omega /c$, where $c$ is the speed of light.
In the absence of the TLS, it simply propagates to the positive direction, $\phi^{(+)}(z,t) = \phi^{(+)}(z-ct,0)$.
At the TLS position, $z_s = 0$, the free field probability amplitude is given by $\phi(0,t) = \phi^{(+)}(-ct,0)$, which exponentially decreases in time.
For this photon pulse, we obtain that 
$\psi(t) = \sqrt{\Gamma_0\Delta/2} \big{(} e^{-(\Gamma_0/2 + i\omega_0)t} - e^{-(\Delta/2 + i\omega_L)t}\big{)} / \big{(}(\Gamma_0-\Delta)/2-i\delta_L\big{)}$. The reactive signatures of the generalized quantum work can now be evidenced.
Firstly, the quantum work by a single photon is an odd function of the detuning, $W_1(-\delta_L) = -W_1(\delta_L)$.
$W_1$ vanishes in the $\delta_L \gg \max\left\{ \Gamma_0, \Delta \right\}$ limit.
Besides, $W_1$ is finite only for finite-time (non-stationary) pulses, since $\lim_{\Delta \rightarrow 0}W_1 = 0$.
Below, we relate this reactive behavior with the real part of the linear susceptibility $\chi'(\omega)$ of the TLS subjected to a weak coherent field. The absorptive signature in the generalized quantum heat is now evidenced.
It can be decomposed as 
$Q_1 = \hbar \omega_0 \big{(} |\psi(t_f)|^2 - |\psi(t_0)|^2 \big{)} + \mathcal{F}[\delta_{\mathrm{eff}}(t)]$,
where 
$\mathcal{F}[\delta_{\mathrm{eff}}(t)] \equiv  \int_{t_0}^{t_f}\big{(} \partial_t |\psi(t)|^2 \big{)} \hbar \delta_{\mathrm{eff}}(t) dt$.
For appreciable values of the TLS excited-state population difference, the term proportional to $\omega_0$ will be much larger then $\mathcal{F}[\delta_{\mathrm{eff}}(t)]$, since we are analyzing here the optical quasi-resonant regime, $\omega_0 \gg \delta_L \sim \delta_{\mathrm{eff}}(t)$.
Below, we relate the generalized quantum heat absorbed from a single photon with the imaginary part of the linear susceptibility $\chi''(\omega)$ of the TLS subjected to a weak coherent field.

{\bf Connexions with a semiclassical scenario.--}
It is worth stressing that the model presented in this paper is not restricted to low TLS excitations.
That is, the TLS can have its population completely inverted, in principle, depending only on the chosen initial single-photon pulse for that.
Nevertheless, the composite TLS-field state always remains in the single-excitation subspace.
This means that some similarities are expected with the low-excitation limit of a TLS driven by a coherent pulse. The purpose of the following discussion is to put the results from the single-photon pulse in perspective with well-known semiclassical results.

A coherent state of the frequency mode $\omega$ is $\ket{\alpha_\omega} = \mathcal{D}[\alpha_\omega] \ket{0_\omega}$, where $\mathcal{D}[\alpha_\omega] = \exp(\alpha_\omega a^\dagger_\omega - \alpha^*_\omega a_\omega)$ is the displacement operator \cite{cohen,scully97}.
A coherent pulse is defined as the tensor product $\ket{\alpha} = \prod_\omega \ket{\alpha_\omega}$.
When each mode is very low populated, $\alpha_\omega \ll 1$, the zero and the one excitation subspaces are predominant, 
$\ket{\alpha \ll 1} \propto \ket{0} + \sum_\omega \alpha_\omega a_\omega^\dagger \ket{0} + \mathcal{O}(\alpha^2)$. 
The single-excitation contribution resembles state $\ket{1}$, in Eq.(\ref{ket1}). When the TLS is driven by a coherent field $\ket{\alpha}$, its reduced state obeys the so-called optical Bloch equations \cite{cohen,scully97}. 
Coherence evolves as
$
\partial_t \rho_{eg}^{(\alpha)} = -(\Gamma_0/2 + i\omega_0)\rho_{eg}^{(\alpha)} - g \alpha(t) \big{(}1 - 2\rho_{ee}^{(\alpha)}\big{)}
$.
The excited-state population obeys 
$\partial_t \rho_{ee}^{(\alpha)} = -\Gamma_0 \rho_{ee}^{(\alpha)} - 2g \mbox{Re}[\alpha(t) (\rho_{eg}^{(\alpha)})^*]$. Here, $\alpha(t) = \sum_\omega \alpha_\omega(0) e^{-i\omega t}$. Note that, in the low-excitation regime, $\rho_{ee}^{(\alpha)}(t) \ll 1$, the equation for the TLS coherence $\rho_{eg}^{(\alpha)}$ is equivalent to Eq.(\ref{psit}), for $\psi(t)$, if $\alpha(t) = \phi(0,t)$. In such a regime, the complex average dipole $d_{\alpha}(t) = d_{eg} \rho_{eg}^{(\alpha)}(t)$  linearly depends on the complex average driving field $E_{\alpha}(t) = i \epsilon_{\omega_0} \alpha(t)$, in the sense that
$d_{\alpha}(t) = \int_{-\infty}^{\infty} \chi(t-t') d_{eg} E_{\alpha}(t') dt' /(\epsilon_{\omega_0} \sqrt{2\pi})$,
where a linear susceptibility, 
$\chi(\tau) = \sqrt{2\pi} \Theta(\tau) i g e^{-(\Gamma_0/2+i\omega_0)\tau}$, 
has been defined.
The frequency domain, 
$\tilde{\chi}(\omega) \equiv \int_{-\infty}^{\infty}  \chi(\tau)e^{i\omega\tau} d\tau/\sqrt{2\pi} = \tilde{\chi}'(\omega)+i\tilde{\chi}''(\omega)$,
provides further physical insight.
The real part, $\tilde{\chi}'(\omega) = g(\omega_0 - \omega)/\big{(}(\Gamma_0/2)^2+(\omega_0-\omega)^2\big{)}$, is at the origin of the field dispersion \cite{scully97}. 
The imaginary part, $\tilde{\chi}''(\omega) = (g\Gamma_0/2)/\big{(}(\Gamma_0/2)^2+(\omega_0-\omega)^2\big{)}$, is related with field damping, which is absorpted by the dipole \cite{cohen,scully97}. Work by a coherent pulse, $W_{\alpha}$, is computed in a full cycle, where the dipole starts and ends at $\ket{g}$, being driven in the meantime by a finite-size pulse
$\alpha(t) = \phi(0,t)$.
Here, the definition of $W_{\alpha}$ from Ref.\cite{cohen} coincides with that from Ref.\cite{alicki79}.
The semiclassical low-excitation (linear) regime is obtained, e.g., in the monochromatic limit, $\Delta \ll \Gamma_0$.
In this case, 
$W_{\alpha} = W^{\mathrm{reac}}_{\alpha} + W^{\mathrm{abs}}_{\alpha}$, in agreement with Ref.\cite{cohen}.
The reactive (or dispersive) contribution is 
$W^{\mathrm{reac}}_{\alpha} =  - \hbar \Delta\ g \int d\omega \tilde{\chi}'(\omega) |\tilde{\alpha}(\omega)|^2$. This is a polarization energy arising when the electric field is in phase with the dipole 
and is classically understood as a reversible form of work stored in the atomic dipole that returns to the field \cite{cohen}.
The absorptive contribution is
$W^{\mathrm{abs}}_{\alpha} = \hbar \omega_L \ 2g \int d\omega \tilde{\chi}''(\omega) |\tilde{\alpha}(\omega)|^2$, 
arising when the electric field is in phase with the time-derivative of the dipole \cite{cohen}.
We defined here that
$\tilde{\alpha}(\omega) 
\equiv \int_{-\infty}^{\infty} \alpha(t') e^{i\omega t'} dt'/\sqrt{2\pi} 
= \sqrt{\varrho_0 \Delta}/\big{(}\Delta/2+i(\omega_L - \omega)\big{)}$.

We find, for the low-excitation regimes of both the quantum model, $|\psi(t)|^2 \ll 1$, and the semiclassical model, $\rho^{(\alpha)}_{ee} \ll 1$, that
\beq
W_{1} 
\approx W^{\mathrm{reac}}_{\alpha}.
\label{W1low}
\eeq
Here, we are using $\alpha(t) = \phi(0,t)$ and $\Delta \ll \Gamma_0$, which imply that 
$\rho^{(\alpha)}_{eg}(t) \approx \psi(t)$ and $\rho^{(\alpha)}_{ee}(t) \approx |\psi(t)|^2$.
More details are given in the Supplemental Material.
We also find that
\beq
Q^{\mathrm{abs}}_{1} 
\approx W^{\mathrm{abs}}_{\alpha},
\label{Q1low}
\eeq
where the generalized quantum heat has been separated in absorptive and emissive contributions,
$Q_1 = Q^{\mathrm{abs}}_{1} + Q^{\mathrm{em}}_{1} $,
being
$Q_{1}^{\mathrm{abs}} \equiv \int_{t_0}^{t_f} \hbar \omega_s(t) \big{(} -2g\mbox{Re}[\phi(0,t) \psi^*(t)]  \big{)} dt$.
Remarkably, Eq.(\ref{Q1low}) means that the amount of generalized heat absorbed from a highly monochromatic single-photon pulse equals the amount of work absorbed from a low-intensity coherent pulse with the same shape.
The fundamental difference is that generalized heat comes from a non-unitary contribution, whereas a coherent pulse provides a unitary contribution to the TLS energy, as is well known in quantum optics \cite{cohen,scully97}. Finally, we find that 
\beq
Q_{1}^{\mathrm{em}}  \approx Q_{\alpha},
\eeq
where $Q_{1}^{\mathrm{em}} \equiv \int_{t_0}^{t_f} 
\big{(} -\Gamma_{0} \hbar\omega_0\  |\psi(t)|^2  
 -\frac{\Gamma_0}{2}  \langle H_{\mathrm{int}} (t) \rangle\big{)} dt$.
Here, $Q_{\alpha}$ is the quantum heat in the semiclassical scenario.
Formally, it is defined in agreement with Ref.\cite{alicki79}.
Physically, it provides the energy lost by spontaneous emission.
More details are given in the Supplemental Material.

{\bf Conclusions.--} 
This paper presents a fully-quantum solution to the problem of quantum work by a single-photon pulse on a TLS.
Quantum work here is defined as the unitary contribution for the TLS energy.
We have shown that this quantum work captures the energetic contribution arising from the dynamical Stark shift in the TLS gap.
The relation between the dynamical Stark shift and the average interaction energy has been elucidated.
We have also shown that a generalized quantum heat captures the energetic contribution arising from the out-of-equilibrium photon absorption and emission by the atom.
This generalized quantum heat is defined here as the non-unitary contribution for the TLS energy coming from the non-Markovian out-of-equilibrium environment due to the interaction with single-photon pulse.
Finally, we have shown that, in the low-excitation regimes of both the quantum and the semiclassical models, the quantum work by a single photon equals the reactive term of the quantum work by a coherent pulse, related with the energy reversibly stored in the dipole.
Still in these low-excitation regimes, the absorbed quantum heat from a single photon was shown to equal the absorbed quantum work by a coherent field.
We emphasized the distinction here that the absorbed quantum heat from a single photon comes as a non-unitary contribution to the TLS reduced dynamics, in contrast with the unitary contribution given by a coherent pulse.

Our results can be regarded as the opening path towards the study of quantum-pulsed machines, since both the shape and the central frequency of the pulse are key variables.
The implications on quantum photocells \cite{scully10}, on quantum optomechanics \cite{yeo13}, and on single-photon heat transport \cite{PRL04,nature06,PRL08,PRL09,PRB12} are promising lines of research.


\

\begin{acknowledgements}

D. V. acknowledges support from CAPES through grant No. 88881.120135/2016-01.
D.V., F.B., T.W. are supported by Instituto Nacional de Ci\^encia e Tecnologia -- Informa\c c\~ao Qu\^antica (INCT-IQ).
\end{acknowledgements}

\section*{References}
\putbib[QuPiston_ArXiv]
\end{bibunit}

\clearpage
\setcounter{footnote}{0}
\setcounter{section}{0}
\setcounter{figure}{0}
\setcounter{table}{0}
\onecolumngrid

\section*{Supplemental Material for ``Quantum work by a single photon''}

The low-excitation regime of the fully-quantum model is defined as $|\psi(t)|^2 \ll 1$.
The low-excitation regime of semiclassical model is $\rho^{(\alpha)}_{ee} \ll 1$.
We choose $\alpha(t) = \phi(0,t)$. 
The highly monochromatic regime is defined as $\Delta \ll \Gamma_0$.
The monochromatic regime implies the low-excitation regime for both the semiclassical and fully-quantum models.
The low-excitation regimes of both the semiclassical and the fully-quantum model imply that 
$\rho^{(\alpha)}_{eg}(t) \approx \psi(t)$, 
$\rho^{(\alpha)}_{ee}(t) \approx |\psi(t)|^2$, 
and
$\langle H_{\mathrm{int},\alpha} (t) \rangle \approx \langle H_{\mathrm{int}} (t) \rangle$ 
(see table below).
The highly monochromatic limit implies that
\beq
- \frac{\partial_t |\rho_{eg}^{(\alpha)}|}{|\rho_{eg}^{(\alpha)}|}  \approx \frac{\Delta}{2}.
\eeq

The reactive work in the low-excitation regime is (see table below)
\beq
W_{\alpha}^{\mathrm{reac}} \approx \frac{\Delta}{2} \int_{t_0}^{t_f} \langle H_{\mathrm{int},\alpha} (t) \rangle dt.
\eeq
The full cycle is defined as $t_0 \rightarrow -\infty$ and $t_f \rightarrow \infty$. 
But
\beq
\int_{-\infty}^{\infty}  \langle H_{\mathrm{int},\alpha}(t) \rangle dt 
= \int_{-\infty}^{\infty} dt \left( -i\hbar g\ \alpha(t) \rho_{eg}^*(t) + c.c. \right)
= -2\hbar g \int d\omega \tilde{\chi}'(\omega) |\tilde{\alpha}(\omega)|^2,
\eeq
where 
$\alpha(t) = \int_{-\infty}^{\infty} \tilde{\alpha}(\omega) e^{-i\omega t} d\omega /\sqrt{2\pi}$,
$\rho_{eg}^{(\alpha)}(t) = \int_{-\infty}^{\infty} \tilde{\rho}_{eg}^{(\alpha)}(\omega) e^{-i\omega t} d\omega /\sqrt{2\pi}$
and
$\rho_{eg}^{(\alpha)}(t) = \int_{-\infty}^{\infty} \chi(t-t') i \alpha(t') dt' /\sqrt{2\pi}$,
where the linear susceptibility is
$\chi(\tau) = \sqrt{2\pi} \Theta(\tau) i g e^{-(\Gamma_0/2+i\omega_0)\tau}$.
In frequency domain,
$\tilde{\rho}_{eg}^{(\alpha)}(\omega) = \tilde{\chi}(\omega)\  i\tilde{\alpha}(\omega)$,
where
$\tilde{\chi}(\omega) 
= \int_{-\infty}^{\infty}  \chi(\tau)e^{i\omega\tau} d\tau/\sqrt{2\pi} 
= \tilde{\chi}'(\omega)+i\tilde{\chi}''(\omega)$.
The real part is $\tilde{\chi}'(\omega) = g(\omega_0 - \omega)/[(\Gamma_0/2)^2+(\omega_0-\omega)^2]$.
The imaginary part is $\tilde{\chi}''(\omega) = (g\Gamma_0/2)/[(\Gamma_0/2)^2+(\omega_0-\omega)^2]$.
For the truncated normalized exponential pulse, $\tilde{\alpha}(\omega) = \sqrt{\varrho_0 \Delta}/[\Delta/2+i(\omega_L - \omega)]$.

For the absorptive contribution (see table below),
\beq
W_{\alpha}^{\mathrm{abs}} 
= \int_{t_0}^{t_f} \hbar \omega_s^{eg}(t) \left( -2g\mbox{Re}[\alpha(t) \rho_{eg}^{(\alpha)*}(t)] \right) dt,
\eeq
we firstly note that, in the highly monochromatic limit, 
\beq
\omega_s^{eg}(t) \approx \omega_s(t) \approx \omega_L.
\eeq
By applying the definition of linear susceptibility given in the above paragraph, it follows that
\beq
\int_{-\infty}^{\infty}  \mbox{Re}[\alpha(t) \rho_{eg}^{(\alpha)*}(t)]  dt 
= - \int d\omega \tilde{\chi}''(\omega)  |\tilde{\alpha}(\omega)|^2.
\eeq
Finally, the interaction energies vanish in the full cycle, $W_{\alpha}^{\mathrm{int}} = W_{1}^{\mathrm{int}} = 0$ (see table below).

\begin{center}
\begin{table*}
\caption{Table of comparisons: semiclassical vs. fully quantum}
    \begin{tabular}{ | l | l  l | p | }

    \hline
   {\bf TLS properties} & {\bf coherent pulse} $\ket{\alpha}$ & {\bf single-photon pulse} $\ket{1}$ \\ \hline
  
             \  & 
             \   & 
             \  \\ \hline

    {\bf Reduced dynamics} & 
    $\dot{\rho}_s = -(i/\hbar)[H_{\alpha}(t),\rho_s] + \mathcal{L}[\rho_s]$ & 
    $\dot{\rho}_s = -(i/\hbar)[H_{1}(t),\rho_s] + \mathcal{L}_t[\rho_s]$  \\ \hline
  
             \  & 
             \   & 
             \  \\ \hline

    {\bf Unitary part} & 
    $H_{\alpha}(t) = H_0 + H_{\mathrm{int},\alpha}(t)$               & 
    $H_{1}(t) = H_0 + H_{\mathrm{int},1}(t) = \hbar\omega_s(t) \ \sigma_{+}\sigma_{-}$  \\ \hline
    
    \   & $H_0 = \hbar\omega_0 \ \sigma_{+}\sigma_{-}$    & $H_0 = \hbar\omega_0 \ \sigma_{+}\sigma_{-}$  \\ \hline
  
      Eff. interaction op. & $H_{\mathrm{int},\alpha} = -i\hbar g\ \alpha(t) \ \sigma_{+} + H.c.$   & $H_{\mathrm{int},1} = \hbar \delta_{\mathrm{eff}}(t) \ \sigma_{+}\sigma_{-}$  \\ \hline
  
         \  &   \ \  & $\omega_s(t) \equiv -\mbox{Im}[\dot{\psi}(t)/\psi(t)] = \omega_0 + \delta_{\mathrm{eff}}(t)$  \\ \hline
     
         \  &  \ \   &$\delta_{\mathrm{eff}}(t) = \langle H_{\mathrm{int}} (t) \rangle/(2|\psi(t)|^2 )$  \\ \hline

         \  & \ \   &  $\langle H_{\mathrm{int}} (t) \rangle = \bra{\xi(t)} H_{\mathrm{int}} \ket{\xi(t)}$  \\ \hline
  
           \  & \  \   & $\ket{\xi(t)} = \psi(t)\ket{e,0} + \sum_\omega \phi_\omega(t) a_\omega^\dagger \ket{g,0}$  \\ \hline
  
             \  & \  \   & $H_{\mathrm{int}} = -i\hbar g \sum_\omega a_\omega \sigma_{+} + H.c.$  \\ \hline   
 
             Interaction energy  & 
             $\langle H_{\mathrm{int},\alpha} (t) \rangle = -i\hbar g\ \alpha(t) \rho_{eg}^*(t) + c.c.$   & 
             $\langle H_{\mathrm{int}} (t) \rangle = -i\hbar g\ \phi(0,t) \psi^*(t) + c.c.$  \\ \hline   

             \  & 
             \   & 
             $\langle H_{\mathrm{int},1} (t) \rangle \equiv \mbox{Tr}[\rho_s H_{\mathrm{int},1}] 
             = \langle H_{\mathrm{int}} (t) \rangle/2$  \\ \hline

             \  & 
             \   & 
             \  \\ \hline

    {\bf Non-unitary part} & 
  $\mathcal{L}[\rho_s]= -\frac{\Gamma_{\mathrm{0}}}{2}(\left\{\sigma_{+}\sigma_{-},\rho_s\right\} -2 \sigma_{-}\rho_s \sigma_{+})$ & 
  $\mathcal{L}_t[\rho_s]= -\frac{\Gamma(t)}{2}(\left\{\sigma_{+}\sigma_{-},\rho_s\right\}-2 \sigma_{-}\rho_s \sigma_{+})$  \\ \hline

\ &   
$\Gamma_{\mathrm{0}} = 4\pi \varrho_{0} g^2$ & 
$\Gamma(t) = \Gamma_{\mathrm{0}} + 2g \mbox{Re}[\phi(0,t)/\psi(t)]$  \\ \hline

           \  & 
             \   & 
             \  \\ \hline

{\bf Matrix elements} & 
$\dot{\rho}_{eg}^{(\alpha)} = -(\frac{\Gamma_{0}}{2} + i\omega_0)\rho_{eg}^{(\alpha)} -g \alpha(t) (1 - 2\rho_{ee}^{(\alpha)})$ & 
$\rho_{eg}^{(1)}(t) = 0$ \\ \hline

\ & $\dot{\rho}_{ee}^{(\alpha)} = - \Gamma_{0} \rho_{ee}^{(\alpha)} -2g \mbox{Re}[\alpha(t) \rho_{eg}^{(\alpha)*}]$ & $\rho_{ee}^{(1)}(t) = |\psi(t)|^2$ \\ \hline

\ &  At low excitations ($\rho_{ee}^{(\alpha)} \ll 1$): & For $0\leq |\psi(t)|^2 \leq 1$, with $\bra{\xi(t)}\xi(t)\rangle = 1$: \\ \hline

\ & $\dot{\rho}_{eg}^{(\alpha)} \approx -(\frac{\Gamma_{0}}{2} + i\omega_0)\rho_{eg}^{(\alpha)} -g \alpha(t)$    & 
$\dot{\psi} = -(\frac{\Gamma_{0}}{2} + i\omega_0)\psi -g \phi(0,t)$  \ \  \\ \hline

           \  & 
             \   & 
             \  \\ \hline   
  
{\bf Work}  & 
             $W_{\alpha} \equiv \int_{t_0}^{t_f} \mbox{Tr}[\rho_s \dot{H}_{\alpha}] dt$   & 
             $W_{1} \equiv \int_{t_0}^{t_f} \mbox{Tr}[\rho_s \dot{H}_{1}] dt$ \\ \hline   


 \ \ & 
$W_{\alpha} = W_{\alpha}^{\mathrm{int}} + W_{\alpha}^{\mathrm{reac}} + W_{\alpha}^{\mathrm{abs}}$
& $W_{1} = W_{1}^{\mathrm{int}} + W_{1}^{\mathrm{reac}}$  \\ \hline

 Interaction energy & 
$W_{\alpha}^{\mathrm{int}} = \langle H_{\mathrm{int},\alpha} (t_f) \rangle - \langle H_{\mathrm{int},\alpha} (t_0) \rangle$ & 
$W_{1}^{\mathrm{int}} = (\langle H_{\mathrm{int}} (t_f) \rangle - \langle H_{\mathrm{int}} (t_0) \rangle)/2$  \\ \hline

 Reactive energy & 
$W_{\alpha}^{\mathrm{reac}} = \int_{t_0}^{t_f} \langle H_{\mathrm{int},\alpha} (t) \rangle \left( - \frac{\partial_t |\rho_{eg}^{(\alpha)}|}{|\rho_{eg}^{(\alpha)}|} \right) dt $ & $W_{1}^{\mathrm{reac}} = \int_{t_0}^{t_f} \langle H_{\mathrm{int}} (t) \rangle  \left( - \frac{\partial_t |\psi|}{|\psi|} \right) dt $  \\ \hline

 Absorbed energy & 
$W_{\alpha}^{\mathrm{abs}} = \int_{t_0}^{t_f} \hbar \omega_s^{eg}(t) \left( -2g\mbox{Re}[\alpha(t) \rho_{eg}^{(\alpha)*}(t)] \right) dt $ \ \  & 
(see $Q_{1}^{\mathrm{abs}}$ below). \\ \hline

       \  &   
       $\omega^{eg}_s(t) \equiv -\mbox{Im}[\dot{\rho}_{eg}^{(\alpha)}/\rho_{eg}^{(\alpha)}]$  & 
       $\omega_s(t) \equiv -\mbox{Im}[\dot{\psi}/\psi]$  \\ \hline

           \  & 
             \   & 
             \  \\ \hline

{\bf Heat}  & 
             $Q_{\alpha} \equiv \int_{t_0}^{t_f} \mbox{Tr}[\dot{\rho}_s {H}_{\alpha}] dt$   & 
             $Q_{1} \equiv \int_{t_0}^{t_f} \mbox{Tr}[\dot{\rho}_s {H}_{1}] dt$ \\ \hline   

\ \   & 
             $Q_{\alpha} = \int_{t_0}^{t_f} \mbox{Tr}[L[{\rho}_s] {H}_{\alpha}] dt$   & 
             $Q_{1} = \int_{t_0}^{t_f} \mbox{Tr}[L_t[{\rho}_s] {H}_{1}] dt$ \\ \hline

 \ \ 
&  \ \ 
& $Q_{1} = Q_{1}^{\mathrm{em}} + Q_{1}^{\mathrm{abs}}$  \\ \hline

 Spont. emitted energy 
& $Q_{\alpha} = -\Gamma_{0} \int_{t_0}^{t_f} \omega_0\   \rho_{ee}^{(\alpha)} \ dt $
& $Q_{1}^{\mathrm{em}} = -\Gamma_{0} \int_{t_0}^{t_f} \omega_0\  |\psi|^2 \ dt $  \\ \hline

& \ \ \ \ \ $+ \left(-\frac{\Gamma_{0}}{2} \int_{t_0}^{t_f} \langle H_{\mathrm{int},\alpha} (t) \rangle dt \right)$
& \ \ \ \ \ \  $+ \left(-\frac{\Gamma_{0}}{2} \int_{t_0}^{t_f} \langle H_{\mathrm{int}} (t) \rangle dt \right)$  \\ \hline

  Absorbed gen. heat 
& (see $W_{\alpha}^{\mathrm{abs}}$ above).
& $Q_{1}^{\mathrm{abs}} = \int_{t_0}^{t_f} \hbar \omega_s(t) \left( -2g\mbox{Re}[\phi(0,t) \psi^*(t)] \right) dt$  \\ \hline

    \end{tabular}
\end{table*}
\end{center}


\begin{thebibliography}{42}%
\makeatletter
\providecommand \@ifxundefined [1]{%
 \@ifx{#1\undefined}
}%
\providecommand \@ifnum [1]{%
 \ifnum #1\expandafter \@firstoftwo
 \else \expandafter \@secondoftwo
 \fi
}%
\providecommand \@ifx [1]{%
 \ifx #1\expandafter \@firstoftwo
 \else \expandafter \@secondoftwo
 \fi
}%
\providecommand \natexlab [1]{#1}%
\providecommand \enquote  [1]{``#1''}%
\providecommand \bibnamefont  [1]{#1}%
\providecommand \bibfnamefont [1]{#1}%
\providecommand \citenamefont [1]{#1}%
\providecommand \href@noop [0]{\@secondoftwo}%
\providecommand \href [0]{\begingroup \@sanitize@url \@href}%
\providecommand \@href[1]{\@@startlink{#1}\@@href}%
\providecommand \@@href[1]{\endgroup#1\@@endlink}%
\providecommand \@sanitize@url [0]{\catcode `\\12\catcode `\$12\catcode
  `\&12\catcode `\#12\catcode `\^12\catcode `\_12\catcode `\%12\relax}%
\providecommand \@@startlink[1]{}%
\providecommand \@@endlink[0]{}%
\providecommand \url  [0]{\begingroup\@sanitize@url \@url }%
\providecommand \@url [1]{\endgroup\@href {#1}{\urlprefix }}%
\providecommand \urlprefix  [0]{URL }%
\providecommand \Eprint [0]{\href }%
\providecommand \doibase [0]{http://dx.doi.org/}%
\providecommand \selectlanguage [0]{\@gobble}%
\providecommand \bibinfo  [0]{\@secondoftwo}%
\providecommand \bibfield  [0]{\@secondoftwo}%
\providecommand \translation [1]{[#1]}%
\providecommand \BibitemOpen [0]{}%
\providecommand \bibitemStop [0]{}%
\providecommand \bibitemNoStop [0]{.\EOS\space}%
\providecommand \EOS [0]{\spacefactor3000\relax}%
\providecommand \BibitemShut  [1]{\csname bibitem#1\endcsname}%
\let\auto@bib@innerbib\@empty
\bibitem [{\citenamefont {Cohen-Tannoudji}\ \emph {et~al.}(1992)\citenamefont
  {Cohen-Tannoudji}, \citenamefont {Dupont-Roc},\ and\ \citenamefont
  {Grynberg}}]{cohen}%
  \BibitemOpen
  \bibfield  {author} {\bibinfo {author} {\bibfnamefont {C.}~\bibnamefont
  {Cohen-Tannoudji}}, \bibinfo {author} {\bibfnamefont {J.}~\bibnamefont
  {Dupont-Roc}}, \ and\ \bibinfo {author} {\bibfnamefont {G.}~\bibnamefont
  {Grynberg}},\ }\href@noop {} {\emph {\bibinfo {title} {Atom-photon
  interactions: basic processes and applications}}}\ (\bibinfo  {publisher}
  {Wiley Online Library},\ \bibinfo {year} {1992})\BibitemShut {NoStop}%
\bibitem [{\citenamefont {Glauber}(1963)}]{glauber63}%
  \BibitemOpen
  \bibfield  {author} {\bibinfo {author} {\bibfnamefont {R.~J.}\ \bibnamefont
  {Glauber}},\ }\href@noop {} {\bibfield  {journal} {\bibinfo  {journal}
  {Physical Review}\ }\textbf {\bibinfo {volume} {130}},\ \bibinfo {pages}
  {2529} (\bibinfo {year} {1963})}\BibitemShut {NoStop}%
\bibitem [{\citenamefont {Domokos}\ \emph {et~al.}(2002)\citenamefont
  {Domokos}, \citenamefont {Horak},\ and\ \citenamefont {Ritsch}}]{domokos}%
  \BibitemOpen
  \bibfield  {author} {\bibinfo {author} {\bibfnamefont {P.}~\bibnamefont
  {Domokos}}, \bibinfo {author} {\bibfnamefont {P.}~\bibnamefont {Horak}}, \
  and\ \bibinfo {author} {\bibfnamefont {H.}~\bibnamefont {Ritsch}},\
  }\href@noop {} {\bibfield  {journal} {\bibinfo  {journal} {Physical Review
  A}\ }\textbf {\bibinfo {volume} {65}},\ \bibinfo {pages} {033832} (\bibinfo
  {year} {2002})}\BibitemShut {NoStop}%
\bibitem [{\citenamefont {Shen}\ and\ \citenamefont {Fan}(2005)}]{fan05}%
  \BibitemOpen
  \bibfield  {author} {\bibinfo {author} {\bibfnamefont {J.~T.}\ \bibnamefont
  {Shen}}\ and\ \bibinfo {author} {\bibfnamefont {S.}~\bibnamefont {Fan}},\
  }\href@noop {} {\bibfield  {journal} {\bibinfo  {journal} {Optics letters}\
  }\textbf {\bibinfo {volume} {30}},\ \bibinfo {pages} {2001} (\bibinfo {year}
  {2005})}\BibitemShut {NoStop}%
\bibitem [{\citenamefont {Chang}\ \emph {et~al.}(2007)\citenamefont {Chang},
  \citenamefont {S{\o}rensen}, \citenamefont {Demler},\ and\ \citenamefont
  {Lukin}}]{chang07}%
  \BibitemOpen
  \bibfield  {author} {\bibinfo {author} {\bibfnamefont {D.~E.}\ \bibnamefont
  {Chang}}, \bibinfo {author} {\bibfnamefont {A.~S.}\ \bibnamefont
  {S{\o}rensen}}, \bibinfo {author} {\bibfnamefont {E.~A.}\ \bibnamefont
  {Demler}}, \ and\ \bibinfo {author} {\bibfnamefont {M.~D.}\ \bibnamefont
  {Lukin}},\ }\href@noop {} {\bibfield  {journal} {\bibinfo  {journal} {Nature
  Phys.}\ }\textbf {\bibinfo {volume} {3}},\ \bibinfo {pages} {807} (\bibinfo
  {year} {2007})}\BibitemShut {NoStop}%
\bibitem [{\citenamefont {Hwang}\ \emph {et~al.}(2009)\citenamefont {Hwang},
  \citenamefont {Pototschnig}, \citenamefont {Lettow}, \citenamefont {Zumofen},
  \citenamefont {Renn}, \citenamefont {G\"otzinger},\ and\ \citenamefont
  {Sandoghdar}}]{sand09}%
  \BibitemOpen
  \bibfield  {author} {\bibinfo {author} {\bibfnamefont {J.}~\bibnamefont
  {Hwang}}, \bibinfo {author} {\bibfnamefont {M.}~\bibnamefont {Pototschnig}},
  \bibinfo {author} {\bibfnamefont {R.}~\bibnamefont {Lettow}}, \bibinfo
  {author} {\bibfnamefont {G.}~\bibnamefont {Zumofen}}, \bibinfo {author}
  {\bibfnamefont {A.}~\bibnamefont {Renn}}, \bibinfo {author} {\bibfnamefont
  {S.}~\bibnamefont {G\"otzinger}}, \ and\ \bibinfo {author} {\bibfnamefont
  {V.}~\bibnamefont {Sandoghdar}},\ }\href@noop {} {\bibfield  {journal}
  {\bibinfo  {journal} {Nature}\ }\textbf {\bibinfo {volume} {460}},\ \bibinfo
  {pages} {76} (\bibinfo {year} {2009})}\BibitemShut {NoStop}%
\bibitem [{\citenamefont {Claudon}\ \emph {et~al.}(2010)\citenamefont
  {Claudon}, \citenamefont {Bleuse}, \citenamefont {Malik}, \citenamefont
  {Bazin}, \citenamefont {Jaffrennou}, \citenamefont {Gregersen}, \citenamefont
  {Sauvan}, \citenamefont {Lalanne},\ and\ \citenamefont {G{\'e}rard}}]{jmg10}%
  \BibitemOpen
  \bibfield  {author} {\bibinfo {author} {\bibfnamefont {J.}~\bibnamefont
  {Claudon}}, \bibinfo {author} {\bibfnamefont {J.}~\bibnamefont {Bleuse}},
  \bibinfo {author} {\bibfnamefont {N.~S.}\ \bibnamefont {Malik}}, \bibinfo
  {author} {\bibfnamefont {M.}~\bibnamefont {Bazin}}, \bibinfo {author}
  {\bibfnamefont {P.}~\bibnamefont {Jaffrennou}}, \bibinfo {author}
  {\bibfnamefont {N.}~\bibnamefont {Gregersen}}, \bibinfo {author}
  {\bibfnamefont {C.}~\bibnamefont {Sauvan}}, \bibinfo {author} {\bibfnamefont
  {P.}~\bibnamefont {Lalanne}}, \ and\ \bibinfo {author} {\bibfnamefont
  {J.-M.}\ \bibnamefont {G{\'e}rard}},\ }\href@noop {} {\bibfield  {journal}
  {\bibinfo  {journal} {Nature Photonics}\ }\textbf {\bibinfo {volume} {4}},\
  \bibinfo {pages} {174} (\bibinfo {year} {2010})}\BibitemShut {NoStop}%
\bibitem [{\citenamefont {Astafiev}\ \emph {et~al.}(2010)\citenamefont
  {Astafiev}, \citenamefont {Abdumalikov~Jr}, \citenamefont {Zagoskin},
  \citenamefont {Pashkin}, \citenamefont {Nakamura},\ and\ \citenamefont
  {Tsai}}]{tsai10}%
  \BibitemOpen
  \bibfield  {author} {\bibinfo {author} {\bibfnamefont {O.}~\bibnamefont
  {Astafiev}}, \bibinfo {author} {\bibfnamefont {A.}~\bibnamefont
  {Abdumalikov~Jr}}, \bibinfo {author} {\bibfnamefont {A.~M.}\ \bibnamefont
  {Zagoskin}}, \bibinfo {author} {\bibfnamefont {Y.~A.}\ \bibnamefont
  {Pashkin}}, \bibinfo {author} {\bibfnamefont {Y.}~\bibnamefont {Nakamura}}, \
  and\ \bibinfo {author} {\bibfnamefont {J.}~\bibnamefont {Tsai}},\ }\href@noop
  {} {\bibfield  {journal} {\bibinfo  {journal} {Physical Review letters}\
  }\textbf {\bibinfo {volume} {104}},\ \bibinfo {pages} {183603} (\bibinfo
  {year} {2010})}\BibitemShut {NoStop}%
\bibitem [{\citenamefont {Scully}(2010)}]{scully10}%
  \BibitemOpen
  \bibfield  {author} {\bibinfo {author} {\bibfnamefont {M.~O.}\ \bibnamefont
  {Scully}},\ }\href@noop {} {\bibfield  {journal} {\bibinfo  {journal}
  {Physical Review letters}\ }\textbf {\bibinfo {volume} {104}},\ \bibinfo
  {pages} {207701} (\bibinfo {year} {2010})}\BibitemShut {NoStop}%
\bibitem [{\citenamefont {Rezus}\ \emph {et~al.}(2012)\citenamefont {Rezus},
  \citenamefont {Walt}, \citenamefont {Lettow}, \citenamefont {Renn},
  \citenamefont {Zumofen}, \citenamefont {G{\"o}tzinger},\ and\ \citenamefont
  {Sandoghdar}}]{sand12}%
  \BibitemOpen
  \bibfield  {author} {\bibinfo {author} {\bibfnamefont {Y.}~\bibnamefont
  {Rezus}}, \bibinfo {author} {\bibfnamefont {S.}~\bibnamefont {Walt}},
  \bibinfo {author} {\bibfnamefont {R.}~\bibnamefont {Lettow}}, \bibinfo
  {author} {\bibfnamefont {A.}~\bibnamefont {Renn}}, \bibinfo {author}
  {\bibfnamefont {G.}~\bibnamefont {Zumofen}}, \bibinfo {author} {\bibfnamefont
  {S.}~\bibnamefont {G{\"o}tzinger}}, \ and\ \bibinfo {author} {\bibfnamefont
  {V.}~\bibnamefont {Sandoghdar}},\ }\href@noop {} {\bibfield  {journal}
  {\bibinfo  {journal} {Physical Review letters}\ }\textbf {\bibinfo {volume}
  {108}},\ \bibinfo {pages} {093601} (\bibinfo {year} {2012})}\BibitemShut
  {NoStop}%
\bibitem [{\citenamefont {Valente}\ \emph {et~al.}(2012)\citenamefont
  {Valente}, \citenamefont {Li}, \citenamefont {Poizat}, \citenamefont
  {G{\'e}rard}, \citenamefont {Kwek}, \citenamefont {Santos},\ and\
  \citenamefont {Auff{\`e}ves}}]{dvNJP}%
  \BibitemOpen
  \bibfield  {author} {\bibinfo {author} {\bibfnamefont {D.}~\bibnamefont
  {Valente}}, \bibinfo {author} {\bibfnamefont {Y.}~\bibnamefont {Li}},
  \bibinfo {author} {\bibfnamefont {J.-P.}\ \bibnamefont {Poizat}}, \bibinfo
  {author} {\bibfnamefont {J.-M.}\ \bibnamefont {G{\'e}rard}}, \bibinfo
  {author} {\bibfnamefont {L.}~\bibnamefont {Kwek}}, \bibinfo {author}
  {\bibfnamefont {M.}~\bibnamefont {Santos}}, \ and\ \bibinfo {author}
  {\bibfnamefont {A.}~\bibnamefont {Auff{\`e}ves}},\ }\href@noop {} {\bibfield
  {journal} {\bibinfo  {journal} {New Journal of Physics}\ }\textbf {\bibinfo
  {volume} {14}},\ \bibinfo {pages} {083029} (\bibinfo {year}
  {2012})}\BibitemShut {NoStop}%
\bibitem [{\citenamefont {Yeo}\ \emph {et~al.}(2014)\citenamefont {Yeo},
  \citenamefont {De~Assis}, \citenamefont {Gloppe}, \citenamefont
  {Dupont-Ferrier}, \citenamefont {Verlot}, \citenamefont {Malik},
  \citenamefont {Dupuy}, \citenamefont {Claudon}, \citenamefont {G{\'e}rard},
  \citenamefont {Auff{\`e}ves} \emph {et~al.}}]{yeo13}%
  \BibitemOpen
  \bibfield  {author} {\bibinfo {author} {\bibfnamefont {I.}~\bibnamefont
  {Yeo}}, \bibinfo {author} {\bibfnamefont {P.-L.}\ \bibnamefont {De~Assis}},
  \bibinfo {author} {\bibfnamefont {A.}~\bibnamefont {Gloppe}}, \bibinfo
  {author} {\bibfnamefont {E.}~\bibnamefont {Dupont-Ferrier}}, \bibinfo
  {author} {\bibfnamefont {P.}~\bibnamefont {Verlot}}, \bibinfo {author}
  {\bibfnamefont {N.~S.}\ \bibnamefont {Malik}}, \bibinfo {author}
  {\bibfnamefont {E.}~\bibnamefont {Dupuy}}, \bibinfo {author} {\bibfnamefont
  {J.}~\bibnamefont {Claudon}}, \bibinfo {author} {\bibfnamefont {J.-M.}\
  \bibnamefont {G{\'e}rard}}, \bibinfo {author} {\bibfnamefont
  {A.}~\bibnamefont {Auff{\`e}ves}},  \emph {et~al.},\ }\href@noop {}
  {\bibfield  {journal} {\bibinfo  {journal} {Nature nanotechnology}\ }\textbf
  {\bibinfo {volume} {9}},\ \bibinfo {pages} {106} (\bibinfo {year}
  {2014})}\BibitemShut {NoStop}%
\bibitem [{\citenamefont {Aljunid}\ \emph {et~al.}(2013)\citenamefont
  {Aljunid}, \citenamefont {Maslennikov}, \citenamefont {Wang}, \citenamefont
  {Dao}, \citenamefont {Scarani},\ and\ \citenamefont {Kurtsiefer}}]{ck}%
  \BibitemOpen
  \bibfield  {author} {\bibinfo {author} {\bibfnamefont {S.~A.}\ \bibnamefont
  {Aljunid}}, \bibinfo {author} {\bibfnamefont {G.}~\bibnamefont
  {Maslennikov}}, \bibinfo {author} {\bibfnamefont {Y.}~\bibnamefont {Wang}},
  \bibinfo {author} {\bibfnamefont {H.~L.}\ \bibnamefont {Dao}}, \bibinfo
  {author} {\bibfnamefont {V.}~\bibnamefont {Scarani}}, \ and\ \bibinfo
  {author} {\bibfnamefont {C.}~\bibnamefont {Kurtsiefer}},\ }\href@noop {}
  {\bibfield  {journal} {\bibinfo  {journal} {Physical Review letters}\
  }\textbf {\bibinfo {volume} {111}},\ \bibinfo {pages} {103001} (\bibinfo
  {year} {2013})}\BibitemShut {NoStop}%
\bibitem [{\citenamefont {Pechal}\ \emph {et~al.}(2014)\citenamefont {Pechal},
  \citenamefont {Huthmacher}, \citenamefont {Eichler}, \citenamefont
  {Zeytino{\u{g}}lu}, \citenamefont {Abdumalikov~Jr}, \citenamefont {Berger},
  \citenamefont {Wallraff},\ and\ \citenamefont {Filipp}}]{wallraff}%
  \BibitemOpen
  \bibfield  {author} {\bibinfo {author} {\bibfnamefont {M.}~\bibnamefont
  {Pechal}}, \bibinfo {author} {\bibfnamefont {L.}~\bibnamefont {Huthmacher}},
  \bibinfo {author} {\bibfnamefont {C.}~\bibnamefont {Eichler}}, \bibinfo
  {author} {\bibfnamefont {S.}~\bibnamefont {Zeytino{\u{g}}lu}}, \bibinfo
  {author} {\bibfnamefont {A.}~\bibnamefont {Abdumalikov~Jr}}, \bibinfo
  {author} {\bibfnamefont {S.}~\bibnamefont {Berger}}, \bibinfo {author}
  {\bibfnamefont {A.}~\bibnamefont {Wallraff}}, \ and\ \bibinfo {author}
  {\bibfnamefont {S.}~\bibnamefont {Filipp}},\ }\href@noop {} {\bibfield
  {journal} {\bibinfo  {journal} {Physical Review X}\ }\textbf {\bibinfo
  {volume} {4}},\ \bibinfo {pages} {041010} (\bibinfo {year}
  {2014})}\BibitemShut {NoStop}%
\bibitem [{\citenamefont {Lodahl}\ \emph {et~al.}(2015)\citenamefont {Lodahl},
  \citenamefont {Mahmoodian},\ and\ \citenamefont {Stobbe}}]{singlephoton}%
  \BibitemOpen
  \bibfield  {author} {\bibinfo {author} {\bibfnamefont {P.}~\bibnamefont
  {Lodahl}}, \bibinfo {author} {\bibfnamefont {S.}~\bibnamefont {Mahmoodian}},
  \ and\ \bibinfo {author} {\bibfnamefont {S.}~\bibnamefont {Stobbe}},\
  }\href@noop {} {\bibfield  {journal} {\bibinfo  {journal} {Reviews of Modern
  Physics}\ }\textbf {\bibinfo {volume} {87}},\ \bibinfo {pages} {347}
  (\bibinfo {year} {2015})}\BibitemShut {NoStop}%
\bibitem [{\citenamefont {Tonner}\ and\ \citenamefont
  {Mahler}(2005)}]{mahler05}%
  \BibitemOpen
  \bibfield  {author} {\bibinfo {author} {\bibfnamefont {F.}~\bibnamefont
  {Tonner}}\ and\ \bibinfo {author} {\bibfnamefont {G.}~\bibnamefont
  {Mahler}},\ }\href@noop {} {\bibfield  {journal} {\bibinfo  {journal}
  {Physical Review E}\ }\textbf {\bibinfo {volume} {72}},\ \bibinfo {pages}
  {066118} (\bibinfo {year} {2005})}\BibitemShut {NoStop}%
\bibitem [{\citenamefont {Weimer}\ \emph {et~al.}(2008)\citenamefont {Weimer},
  \citenamefont {Henrich}, \citenamefont {Rempp}, \citenamefont
  {Schr{\"o}der},\ and\ \citenamefont {Mahler}}]{mahler08}%
  \BibitemOpen
  \bibfield  {author} {\bibinfo {author} {\bibfnamefont {H.}~\bibnamefont
  {Weimer}}, \bibinfo {author} {\bibfnamefont {M.~J.}\ \bibnamefont {Henrich}},
  \bibinfo {author} {\bibfnamefont {F.}~\bibnamefont {Rempp}}, \bibinfo
  {author} {\bibfnamefont {H.}~\bibnamefont {Schr{\"o}der}}, \ and\ \bibinfo
  {author} {\bibfnamefont {G.}~\bibnamefont {Mahler}},\ }\href@noop {}
  {\bibfield  {journal} {\bibinfo  {journal} {EPL (Europhysics Letters)}\
  }\textbf {\bibinfo {volume} {83}},\ \bibinfo {pages} {30008} (\bibinfo {year}
  {2008})}\BibitemShut {NoStop}%
\bibitem [{\citenamefont {Youssef}\ \emph {et~al.}(2009)\citenamefont
  {Youssef}, \citenamefont {Mahler},\ and\ \citenamefont {Obada}}]{QA0}%
  \BibitemOpen
  \bibfield  {author} {\bibinfo {author} {\bibfnamefont {M.}~\bibnamefont
  {Youssef}}, \bibinfo {author} {\bibfnamefont {G.}~\bibnamefont {Mahler}}, \
  and\ \bibinfo {author} {\bibfnamefont {A.-S.}\ \bibnamefont {Obada}},\
  }\href@noop {} {\bibfield  {journal} {\bibinfo  {journal} {Physical Review
  E}\ }\textbf {\bibinfo {volume} {80}},\ \bibinfo {pages} {061129} (\bibinfo
  {year} {2009})}\BibitemShut {NoStop}%
\bibitem [{\citenamefont {Schr{\"o}der}\ and\ \citenamefont
  {Mahler}(2010)}]{QA1}%
  \BibitemOpen
  \bibfield  {author} {\bibinfo {author} {\bibfnamefont {H.}~\bibnamefont
  {Schr{\"o}der}}\ and\ \bibinfo {author} {\bibfnamefont {G.}~\bibnamefont
  {Mahler}},\ }\href@noop {} {\bibfield  {journal} {\bibinfo  {journal}
  {Physical Review E}\ }\textbf {\bibinfo {volume} {81}},\ \bibinfo {pages}
  {021118} (\bibinfo {year} {2010})}\BibitemShut {NoStop}%
\bibitem [{\citenamefont {Xu}(2016)}]{QA3}%
  \BibitemOpen
  \bibfield  {author} {\bibinfo {author} {\bibfnamefont {Y.}~\bibnamefont
  {Xu}},\ }\href@noop {} {\bibfield  {journal} {\bibinfo  {journal} {Physical
  Review E}\ }\textbf {\bibinfo {volume} {94}},\ \bibinfo {pages} {062145}
  (\bibinfo {year} {2016})}\BibitemShut {NoStop}%
\bibitem [{\citenamefont {Alipour}\ \emph {et~al.}(2016)\citenamefont
  {Alipour}, \citenamefont {Benatti}, \citenamefont {Bakhshinezhad},
  \citenamefont {Afsary}, \citenamefont {Marcantoni},\ and\ \citenamefont
  {Rezakhani}}]{QA4}%
  \BibitemOpen
  \bibfield  {author} {\bibinfo {author} {\bibfnamefont {S.}~\bibnamefont
  {Alipour}}, \bibinfo {author} {\bibfnamefont {F.}~\bibnamefont {Benatti}},
  \bibinfo {author} {\bibfnamefont {F.}~\bibnamefont {Bakhshinezhad}}, \bibinfo
  {author} {\bibfnamefont {M.}~\bibnamefont {Afsary}}, \bibinfo {author}
  {\bibfnamefont {S.}~\bibnamefont {Marcantoni}}, \ and\ \bibinfo {author}
  {\bibfnamefont {A.}~\bibnamefont {Rezakhani}},\ }\href@noop {} {\bibfield
  {journal} {\bibinfo  {journal} {Scientific reports}\ }\textbf {\bibinfo
  {volume} {6}} (\bibinfo {year} {2016})}\BibitemShut {NoStop}%
\bibitem [{\citenamefont {Gelbwaser-Klimovsky}\ \emph
  {et~al.}(2013)\citenamefont {Gelbwaser-Klimovsky}, \citenamefont {Alicki},\
  and\ \citenamefont {Kurizki}}]{QAKurizkiAlicki}%
  \BibitemOpen
  \bibfield  {author} {\bibinfo {author} {\bibfnamefont {D.}~\bibnamefont
  {Gelbwaser-Klimovsky}}, \bibinfo {author} {\bibfnamefont {R.}~\bibnamefont
  {Alicki}}, \ and\ \bibinfo {author} {\bibfnamefont {G.}~\bibnamefont
  {Kurizki}},\ }\href@noop {} {\bibfield  {journal} {\bibinfo  {journal} {EPL
  (Europhysics Letters)}\ }\textbf {\bibinfo {volume} {103}},\ \bibinfo {pages}
  {60005} (\bibinfo {year} {2013})}\BibitemShut {NoStop}%
\bibitem [{\citenamefont {Salmilehto}\ \emph {et~al.}(2014)\citenamefont
  {Salmilehto}, \citenamefont {Solinas},\ and\ \citenamefont
  {M{\"o}tt{\"o}nen}}]{QA2}%
  \BibitemOpen
  \bibfield  {author} {\bibinfo {author} {\bibfnamefont {J.}~\bibnamefont
  {Salmilehto}}, \bibinfo {author} {\bibfnamefont {P.}~\bibnamefont {Solinas}},
  \ and\ \bibinfo {author} {\bibfnamefont {M.}~\bibnamefont
  {M{\"o}tt{\"o}nen}},\ }\href@noop {} {\bibfield  {journal} {\bibinfo
  {journal} {Physical Review E}\ }\textbf {\bibinfo {volume} {89}},\ \bibinfo
  {pages} {052128} (\bibinfo {year} {2014})}\BibitemShut {NoStop}%
\bibitem [{\citenamefont {Gelbwaser-Klimovsky}\ \emph
  {et~al.}(2015)\citenamefont {Gelbwaser-Klimovsky}, \citenamefont {Niedenzu},\
  and\ \citenamefont {Kurizki}}]{QA50}%
  \BibitemOpen
  \bibfield  {author} {\bibinfo {author} {\bibfnamefont {D.}~\bibnamefont
  {Gelbwaser-Klimovsky}}, \bibinfo {author} {\bibfnamefont {W.}~\bibnamefont
  {Niedenzu}}, \ and\ \bibinfo {author} {\bibfnamefont {G.}~\bibnamefont
  {Kurizki}},\ }\href@noop {} {\bibfield  {journal} {\bibinfo  {journal}
  {Advances In Atomic, Molecular, and Optical Physics}\ }\textbf {\bibinfo
  {volume} {64}},\ \bibinfo {pages} {329} (\bibinfo {year} {2015})}\BibitemShut
  {NoStop}%
\bibitem [{\citenamefont {Mitchison}\ \emph {et~al.}(2016)\citenamefont
  {Mitchison}, \citenamefont {Huber}, \citenamefont {Prior}, \citenamefont
  {Woods},\ and\ \citenamefont {Plenio}}]{QA5}%
  \BibitemOpen
  \bibfield  {author} {\bibinfo {author} {\bibfnamefont {M.~T.}\ \bibnamefont
  {Mitchison}}, \bibinfo {author} {\bibfnamefont {M.}~\bibnamefont {Huber}},
  \bibinfo {author} {\bibfnamefont {J.}~\bibnamefont {Prior}}, \bibinfo
  {author} {\bibfnamefont {M.~P.}\ \bibnamefont {Woods}}, \ and\ \bibinfo
  {author} {\bibfnamefont {M.~B.}\ \bibnamefont {Plenio}},\ }\href@noop {}
  {\bibfield  {journal} {\bibinfo  {journal} {Quantum Science and Technology}\
  }\textbf {\bibinfo {volume} {1}},\ \bibinfo {pages} {015001} (\bibinfo {year}
  {2016})}\BibitemShut {NoStop}%
\bibitem [{\citenamefont {Hofer}\ \emph {et~al.}(2016)\citenamefont {Hofer},
  \citenamefont {Perarnau-Llobet}, \citenamefont {Brask}, \citenamefont
  {Silva}, \citenamefont {Huber},\ and\ \citenamefont {Brunner}}]{QA6}%
  \BibitemOpen
  \bibfield  {author} {\bibinfo {author} {\bibfnamefont {P.~P.}\ \bibnamefont
  {Hofer}}, \bibinfo {author} {\bibfnamefont {M.}~\bibnamefont
  {Perarnau-Llobet}}, \bibinfo {author} {\bibfnamefont {J.~B.}\ \bibnamefont
  {Brask}}, \bibinfo {author} {\bibfnamefont {R.}~\bibnamefont {Silva}},
  \bibinfo {author} {\bibfnamefont {M.}~\bibnamefont {Huber}}, \ and\ \bibinfo
  {author} {\bibfnamefont {N.}~\bibnamefont {Brunner}},\ }\href@noop {}
  {\bibfield  {journal} {\bibinfo  {journal} {Physical Review B}\ }\textbf
  {\bibinfo {volume} {94}},\ \bibinfo {pages} {235420} (\bibinfo {year}
  {2016})}\BibitemShut {NoStop}%
\bibitem [{\citenamefont {Roulet}\ \emph {et~al.}(2017)\citenamefont {Roulet},
  \citenamefont {Nimmrichter}, \citenamefont {Arrazola}, \citenamefont {Seah},\
  and\ \citenamefont {Scarani}}]{QA7}%
  \BibitemOpen
  \bibfield  {author} {\bibinfo {author} {\bibfnamefont {A.}~\bibnamefont
  {Roulet}}, \bibinfo {author} {\bibfnamefont {S.}~\bibnamefont {Nimmrichter}},
  \bibinfo {author} {\bibfnamefont {J.~M.}\ \bibnamefont {Arrazola}}, \bibinfo
  {author} {\bibfnamefont {S.}~\bibnamefont {Seah}}, \ and\ \bibinfo {author}
  {\bibfnamefont {V.}~\bibnamefont {Scarani}},\ }\href@noop {} {\bibfield
  {journal} {\bibinfo  {journal} {Physical Review E}\ }\textbf {\bibinfo
  {volume} {95}},\ \bibinfo {pages} {062131} (\bibinfo {year}
  {2017})}\BibitemShut {NoStop}%
\bibitem [{\citenamefont {Schmidt}\ \emph {et~al.}(2004)\citenamefont
  {Schmidt}, \citenamefont {Schoelkopf},\ and\ \citenamefont
  {Cleland}}]{PRL04}%
  \BibitemOpen
  \bibfield  {author} {\bibinfo {author} {\bibfnamefont {D.}~\bibnamefont
  {Schmidt}}, \bibinfo {author} {\bibfnamefont {R.}~\bibnamefont {Schoelkopf}},
  \ and\ \bibinfo {author} {\bibfnamefont {A.}~\bibnamefont {Cleland}},\
  }\href@noop {} {\bibfield  {journal} {\bibinfo  {journal} {Physical Review
  letters}\ }\textbf {\bibinfo {volume} {93}},\ \bibinfo {pages} {045901}
  (\bibinfo {year} {2004})}\BibitemShut {NoStop}%
\bibitem [{\citenamefont {Meschke}\ \emph {et~al.}(2006)\citenamefont
  {Meschke}, \citenamefont {Guichard},\ and\ \citenamefont
  {Pekola}}]{nature06}%
  \BibitemOpen
  \bibfield  {author} {\bibinfo {author} {\bibfnamefont {M.}~\bibnamefont
  {Meschke}}, \bibinfo {author} {\bibfnamefont {W.}~\bibnamefont {Guichard}}, \
  and\ \bibinfo {author} {\bibfnamefont {J.~P.}\ \bibnamefont {Pekola}},\
  }\href@noop {} {\bibfield  {journal} {\bibinfo  {journal} {Nature}\ }\textbf
  {\bibinfo {volume} {444}},\ \bibinfo {pages} {187} (\bibinfo {year}
  {2006})}\BibitemShut {NoStop}%
\bibitem [{\citenamefont {Ojanen}\ and\ \citenamefont {Jauho}(2008)}]{PRL08}%
  \BibitemOpen
  \bibfield  {author} {\bibinfo {author} {\bibfnamefont {T.}~\bibnamefont
  {Ojanen}}\ and\ \bibinfo {author} {\bibfnamefont {A.-P.}\ \bibnamefont
  {Jauho}},\ }\href@noop {} {\bibfield  {journal} {\bibinfo  {journal}
  {Physical Review letters}\ }\textbf {\bibinfo {volume} {100}},\ \bibinfo
  {pages} {155902} (\bibinfo {year} {2008})}\BibitemShut {NoStop}%
\bibitem [{\citenamefont {Timofeev}\ \emph {et~al.}(2009)\citenamefont
  {Timofeev}, \citenamefont {Helle}, \citenamefont {Meschke}, \citenamefont
  {M{\"o}tt{\"o}nen},\ and\ \citenamefont {Pekola}}]{PRL09}%
  \BibitemOpen
  \bibfield  {author} {\bibinfo {author} {\bibfnamefont {A.~V.}\ \bibnamefont
  {Timofeev}}, \bibinfo {author} {\bibfnamefont {M.}~\bibnamefont {Helle}},
  \bibinfo {author} {\bibfnamefont {M.}~\bibnamefont {Meschke}}, \bibinfo
  {author} {\bibfnamefont {M.}~\bibnamefont {M{\"o}tt{\"o}nen}}, \ and\
  \bibinfo {author} {\bibfnamefont {J.~P.}\ \bibnamefont {Pekola}},\
  }\href@noop {} {\bibfield  {journal} {\bibinfo  {journal} {Physical Review
  letters}\ }\textbf {\bibinfo {volume} {102}},\ \bibinfo {pages} {200801}
  (\bibinfo {year} {2009})}\BibitemShut {NoStop}%
\bibitem [{\citenamefont {Jones}\ \emph {et~al.}(2012)\citenamefont {Jones},
  \citenamefont {Huhtam{\"a}ki}, \citenamefont {Tan},\ and\ \citenamefont
  {M{\"o}tt{\"o}nen}}]{PRB12}%
  \BibitemOpen
  \bibfield  {author} {\bibinfo {author} {\bibfnamefont {P.}~\bibnamefont
  {Jones}}, \bibinfo {author} {\bibfnamefont {J.}~\bibnamefont
  {Huhtam{\"a}ki}}, \bibinfo {author} {\bibfnamefont {K.}~\bibnamefont {Tan}},
  \ and\ \bibinfo {author} {\bibfnamefont {M.}~\bibnamefont
  {M{\"o}tt{\"o}nen}},\ }\href@noop {} {\bibfield  {journal} {\bibinfo
  {journal} {Physical Review B}\ }\textbf {\bibinfo {volume} {85}},\ \bibinfo
  {pages} {075413} (\bibinfo {year} {2012})}\BibitemShut {NoStop}%
\bibitem [{\citenamefont {Hekking}\ and\ \citenamefont
  {Pekola}(2013)}]{PhotonOutThEq}%
  \BibitemOpen
  \bibfield  {author} {\bibinfo {author} {\bibfnamefont {F.}~\bibnamefont
  {Hekking}}\ and\ \bibinfo {author} {\bibfnamefont {J.~P.}\ \bibnamefont
  {Pekola}},\ }\href@noop {} {\bibfield  {journal} {\bibinfo  {journal}
  {Physical Review letters}\ }\textbf {\bibinfo {volume} {111}},\ \bibinfo
  {pages} {093602} (\bibinfo {year} {2013})}\BibitemShut {NoStop}%
\bibitem [{\citenamefont {Schmidt}\ \emph {et~al.}(2015)\citenamefont
  {Schmidt}, \citenamefont {Carusela}, \citenamefont {Pekola}, \citenamefont
  {Suomela},\ and\ \citenamefont {Ankerhold}}]{NMQuTh}%
  \BibitemOpen
  \bibfield  {author} {\bibinfo {author} {\bibfnamefont {R.}~\bibnamefont
  {Schmidt}}, \bibinfo {author} {\bibfnamefont {M.~F.}\ \bibnamefont
  {Carusela}}, \bibinfo {author} {\bibfnamefont {J.~P.}\ \bibnamefont
  {Pekola}}, \bibinfo {author} {\bibfnamefont {S.}~\bibnamefont {Suomela}}, \
  and\ \bibinfo {author} {\bibfnamefont {J.}~\bibnamefont {Ankerhold}},\
  }\href@noop {} {\bibfield  {journal} {\bibinfo  {journal} {Physical Review
  B}\ }\textbf {\bibinfo {volume} {91}},\ \bibinfo {pages} {224303} (\bibinfo
  {year} {2015})}\BibitemShut {NoStop}%
\bibitem [{\citenamefont {Pusz}\ and\ \citenamefont
  {Woronowicz}(1978)}]{pusz78}%
  \BibitemOpen
  \bibfield  {author} {\bibinfo {author} {\bibfnamefont {W.}~\bibnamefont
  {Pusz}}\ and\ \bibinfo {author} {\bibfnamefont {S.}~\bibnamefont
  {Woronowicz}},\ }\href@noop {} {\bibfield  {journal} {\bibinfo  {journal}
  {Communications in Mathematical Physics}\ }\textbf {\bibinfo {volume} {58}},\
  \bibinfo {pages} {273} (\bibinfo {year} {1978})}\BibitemShut {NoStop}%
\bibitem [{\citenamefont {Alicki}(1979)}]{alicki79}%
  \BibitemOpen
  \bibfield  {author} {\bibinfo {author} {\bibfnamefont {R.}~\bibnamefont
  {Alicki}},\ }\href@noop {} {\bibfield  {journal} {\bibinfo  {journal}
  {Journal of Physics A: Mathematical and General}\ }\textbf {\bibinfo {volume}
  {12}},\ \bibinfo {pages} {L103} (\bibinfo {year} {1979})}\BibitemShut
  {NoStop}%
\bibitem [{\citenamefont {Greiner}(2009)}]{classmech19351}%
  \BibitemOpen
  \bibfield  {author} {\bibinfo {author} {\bibfnamefont {W.}~\bibnamefont
  {Greiner}},\ }\href@noop {} {\emph {\bibinfo {title} {Classical mechanics:
  systems of particles and Hamiltonian dynamics}}}\ (\bibinfo  {publisher}
  {Springer Science \& Business Media},\ \bibinfo {year} {2009})\BibitemShut
  {NoStop}%
\bibitem [{\citenamefont {Arnol'd}(2013)}]{classmech19352}%
  \BibitemOpen
  \bibfield  {author} {\bibinfo {author} {\bibfnamefont {V.~I.}\ \bibnamefont
  {Arnol'd}},\ }\href@noop {} {\emph {\bibinfo {title} {Mathematical methods of
  classical mechanics}}}\ (\bibinfo  {publisher} {Springer Science \& Business
  Media},\ \bibinfo {year} {2013})\BibitemShut {NoStop}%
\bibitem [{\citenamefont {Drobn{\`y}}\ \emph {et~al.}(2000)\citenamefont
  {Drobn{\`y}}, \citenamefont {Havukainen},\ and\ \citenamefont
  {Buzek}}]{buzek00}%
  \BibitemOpen
  \bibfield  {author} {\bibinfo {author} {\bibfnamefont {G.}~\bibnamefont
  {Drobn{\`y}}}, \bibinfo {author} {\bibfnamefont {M.}~\bibnamefont
  {Havukainen}}, \ and\ \bibinfo {author} {\bibfnamefont {V.}~\bibnamefont
  {Buzek}},\ }\href@noop {} {\bibfield  {journal} {\bibinfo  {journal} {Journal
  of Modern Optics}\ }\textbf {\bibinfo {volume} {47}},\ \bibinfo {pages} {851}
  (\bibinfo {year} {2000})}\BibitemShut {NoStop}%
\bibitem [{\citenamefont {Valente}\ \emph {et~al.}(2016)\citenamefont
  {Valente}, \citenamefont {Arruda},\ and\ \citenamefont {Werlang}}]{dv16}%
  \BibitemOpen
  \bibfield  {author} {\bibinfo {author} {\bibfnamefont {D.}~\bibnamefont
  {Valente}}, \bibinfo {author} {\bibfnamefont {M.}~\bibnamefont {Arruda}}, \
  and\ \bibinfo {author} {\bibfnamefont {T.}~\bibnamefont {Werlang}},\
  }\href@noop {} {\bibfield  {journal} {\bibinfo  {journal} {Optics letters}\
  }\textbf {\bibinfo {volume} {41}},\ \bibinfo {pages} {3126} (\bibinfo {year}
  {2016})}\BibitemShut {NoStop}%
\bibitem [{\citenamefont {Valente}\ \emph {et~al.}(2017)\citenamefont
  {Valente}, \citenamefont {Brito},\ and\ \citenamefont {Werlang}}]{dv17}%
  \BibitemOpen
  \bibfield  {author} {\bibinfo {author} {\bibfnamefont {D.}~\bibnamefont
  {Valente}}, \bibinfo {author} {\bibfnamefont {F.}~\bibnamefont {Brito}}, \
  and\ \bibinfo {author} {\bibfnamefont {T.}~\bibnamefont {Werlang}},\
  }\href@noop {} {\bibfield  {journal} {\bibinfo  {journal} {Optics Letters}\
  }\textbf {\bibinfo {volume} {42}},\ \bibinfo {pages} {1692} (\bibinfo {year}
  {2017})}\BibitemShut {NoStop}%
\bibitem [{\citenamefont {Scully}\ and\ \citenamefont
  {Zubairy}(1997)}]{scully97}%
  \BibitemOpen
  \bibfield  {author} {\bibinfo {author} {\bibfnamefont {M.~O.}\ \bibnamefont
  {Scully}}\ and\ \bibinfo {author} {\bibfnamefont {M.~S.}\ \bibnamefont
  {Zubairy}},\ }\href@noop {} {\emph {\bibinfo {title} {Quantum Optics}}}\
  (\bibinfo  {publisher} {Cambridge University Press},\ \bibinfo {year}
  {1997})\BibitemShut {NoStop}%
\end{thebibliography}%
\end{document}